%% file: root.tex
\documentclass[letterpaper, 10 pt, conference]{ieeeconf}  % Comment this line out if you need a4paper

\IEEEoverridecommandlockouts                              % This command is only needed if 
\usepackage{amsmath}
\usepackage{amssymb}
\usepackage{graphicx}
\usepackage{upgreek}
\usepackage{algorithm}
\usepackage{algpseudocode}
\usepackage{verbatim}

\usepackage{mcode}
\usepackage{booktabs}
\usepackage{units}
\usepackage{subfig}

\usepackage{tikz}
\usepackage{pgfplots}
\pgfplotsset{compat=newest}
\usepackage{pgfkeys}
\newlength\myheight
\newlength\mywidth
\usetikzlibrary{plotmarks}
\usetikzlibrary{shapes,arrows,shadows,positioning}
\usepackage{tikzscale}
\usetikzlibrary{decorations.pathreplacing}
\usetikzlibrary{shapes,arrows,patterns}
\newtheorem{theorem}{Theorem}
\newtheorem{corollary}{Corollary}

\newtheorem{lemma}{Lemma}
\newtheorem{definition}{Definition}
\newtheorem{remark}{Remark}

\input{auxfiles/blockschaltbilder}

\title{\LARGE \bf
Practicable Simulation-Free Model Order Reduction \\ by Nonlinear Moment Matching
}

\author{M. Cruz Varona, R. Gebhart, J. Suk and B. Lohmann% <-this % stops a space
	\thanks{M. Cruz Varona is with the Chair of Automatic Control, Department of Mechanical Engineering, Technical University of Munich, Boltzmannstr. 15, 85748 Garching, Germany. 
		E-mail: {\tt\small maria.cruz@tum.de}}%
}

\begin{document}

\maketitle
\thispagestyle{empty}
\pagestyle{empty}

%%%%%%%%%%%%%%%%%%%%%%%%%%%%%%%%%%%%%%%%%%%%%%%%%%%%%%%%%%%%%%%%%%%%%%%%%%%%%%%%
\begin{abstract}
	In this paper, a practicable simulation-free model order reduction method by nonlinear moment matching is developed. Based on the steady-state interpretation of linear moment matching, we comprehensively explain the extension of this reduction concept to nonlinear systems presented in~\cite{astolfi2010model}, provide some new insights and propose some simplifications to achieve a feasible and numerically efficient nonlinear model reduction algorithm. This algorithm relies on the solution of nonlinear systems of equations rather than on the expensive simulation of the original model or the difficult solution of a nonlinear partial differential equation. %, as employed in standard nonlinear approaches.  
\end{abstract}

%%%%%%%%%%%%%%%%%%%%%%%%%%%%%%%%%%%%%%%%%%%%%%%%%%%%%%%%%%%%%%%%%%%%%%%%%%%%%%%%
\vspace{-0.2em}
\section{INTRODUCTION}

%of very large dimension
%Nonlinear model order reduction has gained a lot of attention over the past two decades
%The detailed mathematical modeling of complex dynamical systems by e.g. the finite element method often yields models of thousands or even millions of degrees of freedom. Since the numerical simulation, optimization and control design for such large-scale systems are computatio\-nally too demanding, reduced order models that accurately approximate the original systems with substantially less effort are strongly aimed. 

The detailed mathematical modeling of complex dynamical systems often yields models of thousands of degrees of freedom. Since the numerical simulation and design for such large-scale systems is computatio\-nally too demanding, reduced order models that accurately approximate the original systems with substantially less effort are strongly aimed.

%The need for simulating large-scale \emph{nonlinear} systems --> Nonlinear model order reduction 

Nonlinear model order reduction has been widely studied over the past two decades due to the ever increasing interest in the efficient, numerical analysis of large-scale \emph{nonlinear} dynamical systems. In this regard, \emph{simulation-based} reduction techniques such as Proper Orthogonal Decomposition (POD) \cite{moore1981principal, kunisch1999control}, Trajectory Piecewise Linear (TPWL) \cite{rewienski2003trajectory}, Empirical Gramians \cite{lall2002subspace}, Balanced-POD \cite{willcox2002balanced} and Reduced Basis methods \cite{haasdonk2017reduced} have established as standard approaches. \emph{System-theoretic} reduction procedures such as nonlinear ba\-lan\-ced truncation \cite{scherpen1993balancing} and Krylov subspace methods for special nonlinear system classes \cite{breiten2013interpolatory} have been also studied. \\ Recently, the concept of moment matching and Krylov subspaces has been developed and carried over to nonlinear systems \cite{astolfi2010model, astolfi2010steady, ionescu2013families, ionescu2016nonlinear, scarciotti2016model, scarciotti2017data}. From a system-theoretical perspective, this represents a promising and interesting a\-pproach towards a nonlinear model reduction technique, which does not rely on the numerical simulation of the original full model to construct the reduced model.

The extension of the well-known concept of moment matching to nonlinear systems has been initiated by Astolfi \cite{astolfi2010model} a few years ago based on the theory of the steady-state res\-ponse of nonlinear systems and the techniques of nonlinear output regulation \cite{krener1992construction}, \cite[Ch. 8]{isidori1995nonlinear}, \cite{huang2004nonlinear}. Since then, moment matching for linear and, specially, nonlinear systems has been further developed in several pu\-bli\-cations. For instance, the equivalence between projection-based and parametrized, non-projective families of reduced models achieving moment matching is presented in \cite{astolfi2010steady}. Therein, the time domain interpretation of \emph{output Krylov subspace}-based moment matching is also established for linear systems using the dual Sylvester equation. These findings are transferred to the nonlinear case in \cite{ionescu2013families} and further developed in \cite{ionescu2016nonlinear} to provide a two-sided, nonlinear moment matching theory. More recently, the steady-state interpretation of moments has also been extended to linear and nonlinear time-delay systems \cite{scarciotti2016model}. In addition, the online estimation of moments of linear and nonlinear systems from input/output data has newly been proposed in \cite{scarciotti2017data}. This recent work aims at the data-driven, low-order identification of an unknown nonlinear system, by solving a recursive, moving window, least-square estimation problem using input/output snapshot measurements. Hereby, a polynomial expansion ansatz with user-defined basis functions is used for the reduced output mapping. Furthermore, the concept of parametrized families of reduced models is employed to enforce additional properties on the reduced order model, which approximately matches the nonlinear, estimated moments. From a practical point of view, \cite{scarciotti2017data} represents the first milestone towards a feasible method for nonlinear moment matching, since the proposed algorithm does not involve the solution of a partial differential equation, but rather aims at estimating the moment of a nonlinear system from input/output data. In fact, the previously mentioned papers \cite{astolfi2010model,ionescu2013families,ionescu2016nonlinear,scarciotti2016model} face the same difficulty, namely the computation of the solution of a nonlinear Sylvester-like partial differential equation.

This paper de\-ve\-lopes the concept of moment matching for nonlinear systems presented in \cite{astolfi2010model} towards practical application: Inspired by the POD community, which usually employs a linear projection and time-snapshots to reduce nonlinear systems, we propose some simplifications to avoid the Sylvester-like partial differential equation and achieve a feasible, numerical algorithm for model reduction by nonlinear moment matching. The proposed approach is actually linked to the technique presented in \cite{scarciotti2017data} in the sense that both methods \emph{approximately} match nonlinear moments. However, the goals of both techniques are different, since \cite{scarciotti2017data} focuses on the data-driven, \emph{low-order identification} of an unknown nonlinear system, whereas this paper deals with the \emph{reduction} of a known nonlinear system. For this reason, the proposed algorithms are also different. Algorithm 2 in \cite{scarciotti2017data} requires the solution of a least-square problem using input/output data, whereas the practicable algorithm presented here relies on the solution of nonlinear systems of equations using the explicitly known nonlinear system.

The rest of the paper is organized as follows. In Section~\ref{sec:LMM} the theory of model reduction by moment matching for linear systems is first reviewed. Herein, the time domain perception of moment matching from \cite{astolfi2010model,astolfi2010steady} as the interpolation of the steady-state response of the output of the system when excited by exponential inputs plays a crucial role for transferring this theory to nonlinear systems. After comprehensively explaining the generalization given in \cite{astolfi2010model} and providing some valuable, illuminating insights in Section \ref{sec:nlmm-PDE}, some step-by-step simplifications are performed in Section~\ref{sec:simulation-free-NLMM} towards a \emph{practicable}, \emph{simulation-free} algorithm for nonlinear moment matching. Finally, a numerical example is provided, which illustrates the effectiveness of the proposed method.

%offers a better understanding, provides new insights, 

%multiple-input multiple-output (MIMO)

\textbf{Notation:} $\mathbb{R}$ is the set of real numbers and $\mathbb{C}$ is the set of complex numbers. $\uplambda(\boldsymbol{E}^{-1} \boldsymbol{A})$ denotes the spectrum of the matrix $\boldsymbol{E}^{-1} \boldsymbol{A} \in \mathbb{R}^{n \times n}$ and $\emptyset$ represents the empty set. Finally, the range of a matrix $\boldsymbol{V}$ is denoted by $\mathrm{Ran}(\boldsymbol{V})$.
%$\mathbb{R}$ is the set of real numbers and $\mathbb{C}$ is the set of complex numbers. 

%%%%%%%%%%%%%%%%%%%%%%%%%%%%%%%%%%%%%%%%%%%%%%%%%%%%%%%%%%%%%%%%%%%%%%%%%%%%%%%%
%\vspace{-0.1em}
\section{Moment Matching for linear systems} \label{sec:LMM}%Implicit moment matching by Krylov-subspace methods %Notion of moments % Krylov subspaces %Sylvester equations 
%\subsection{Linear model reduction by projection}
Consider a large-scale, linear time-invariant (LTI), asymptotically stable, multiple-input multiple-output (MIMO) state-space model of the form
\vspace{-0.2em}
\begin{equation} \label{eq:linear-FOM}
	\boldsymbol{E} \, \dot{\boldsymbol{x}}(t) = \boldsymbol{A} \, \boldsymbol{x}(t) + \boldsymbol{B} \, \boldsymbol{u}(t), \quad \boldsymbol{y}(t) = \boldsymbol{C} \, \boldsymbol{x}(t),
\end{equation}
%\vspace{-0.1em}
where $\boldsymbol{E} \in \mathbb{R}^{n \times n}$ with $\det(\boldsymbol{E}) \neq 0$ is the descriptor matrix, $\boldsymbol{A} \in \mathbb{R}^{n \times n}$ is the system matrix and $\boldsymbol{x}(t) \in \mathbb{R}^n$, $\boldsymbol{u}(t) \in \mathbb{R}^m$, $\boldsymbol{y}(t) \in \mathbb{R}^p$ denote the state, inputs and outputs of the system, respectively. The input-output behavior is characterized in the frequency domain by the rational transfer function
\vspace{-0.2em}
\begin{equation}
	\boldsymbol{G}(s) = \boldsymbol{C}(s \boldsymbol{E} - \boldsymbol{A})^{-1} \boldsymbol{B} \ \ \in \, \mathbb{C}^{p \times m}.
\end{equation}
%\vspace{-0.1em}
The goal of model order reduction is to approximate the full order model (FOM) \eqref{eq:linear-FOM} by a reduced order model (ROM)
\vspace{-0.2em}
\begin{equation} \label{eq:linear-ROM}
	\boldsymbol{E}_{\mathrm{r}} \, \dot{\boldsymbol{x}}_{\mathrm{r}}(t) = \boldsymbol{A}_{\mathrm{r}} \, \boldsymbol{x}_{\mathrm{r}}(t) + \boldsymbol{B}_{\mathrm{r}} \, \boldsymbol{u}(t), \quad \boldsymbol{y}_{\mathrm{r}}(t) = \boldsymbol{C}_{\mathrm{r}} \, \boldsymbol{x}_{\mathrm{r}}(t),
\end{equation}
%\vspace{-0.1em}
of much smaller dimension $r \ll n$ with $\boldsymbol{E}_{\mathrm{r}} \!=\! \boldsymbol{W}^{\mathsf{T}} \boldsymbol{E} \boldsymbol{V}$, $\boldsymbol{A}_{\mathrm{r}} \!=\! \boldsymbol{W}^{\mathsf{T}} \boldsymbol{A} \boldsymbol{V}$, $\boldsymbol{B}_{\mathrm{r}} \!=\! \boldsymbol{W}^{\mathsf{T}} \boldsymbol{B}$ and $\boldsymbol{C}_{\mathrm{r}} \!=\! \boldsymbol{C} \boldsymbol{V}$, such that $\boldsymbol{y}(t) \approx \boldsymbol{y}_{\mathrm{r}}(t)$. Note that in this framework, the reduction is performed by a \emph{Petrov-Galerkin projection} of \eqref{eq:linear-FOM} onto the $r$-dimensional subspace $\mathrm{Ran}(\boldsymbol{EV})$ by means of the projector $\boldsymbol{P} \!=\! \boldsymbol{E} \boldsymbol{V}(\boldsymbol{W}^{\mathsf T} \boldsymbol{E} \boldsymbol{V})^{-1} \boldsymbol{W}^{\mathsf T}$. Thus, the main task in this setting consists in finding suitable (orthogonal) projection matrices $\boldsymbol{V}, \boldsymbol{W} \in \mathbb{R}^{n \times r}$ that span appropriate subspaces. %such that $\det(\boldsymbol{W}^{\mathsf T} \boldsymbol{E} \boldsymbol{V}) \neq 0$.

%=================================
%$\det(\boldsymbol{W}^{\mathsf T} \boldsymbol{E} \boldsymbol{V}) \neq 0$
%\vspace{-0.5em}
\subsection{Notion of Moments and Krylov subspaces}
One common and numerically efficient linear reduction technique relies on the concept of \emph{implicit moment matching} by rational Krylov subspaces \cite{grimme1997krylov, beattie2017model}. 
\begin{definition}
	The Taylor series expansion of the transfer function $\boldsymbol{G}(s)$ around a complex number $\sigma \in \mathbb{C}$, also called \emph{shift} or \emph{expansion / interpolation point}, is given by
	\vspace{-0.2em}
	\begin{equation}
		\boldsymbol{G}(s) = \sum_{i=0}^{\infty} \boldsymbol{m}_{i}(\sigma) (s - \sigma)^i\,,
	\end{equation}
	where $\boldsymbol{m}_i(\sigma)$ is called the $i$-th \emph{moment} of $\boldsymbol{G}(s)$ around $\sigma$. The moments represent the Taylor coefficients and satisfy:
	\begin{equation*}
		\begin{aligned}
			\boldsymbol{m}_i(\sigma) &= \frac{1}{i!} \left.\frac{\mathrm{d}^{i} \boldsymbol{G}(s)}{\mathrm{d}s^{i}}\right|_{s=\sigma} = \frac{1}{i!} \left.\left[\frac{\mathrm{d}^{i}}{\mathrm{d}s^{i}} \, \boldsymbol{C}(s\boldsymbol{E} \!-\! \boldsymbol{A})^{-1}\boldsymbol{B}\right]\right|_{s=\sigma} \\[0.5em]
			&=(-1)^i \, \boldsymbol{C} \big((\sigma \boldsymbol{E} - \boldsymbol{A})^{-1} \boldsymbol{E}\big)^i (\sigma \boldsymbol{E} - \boldsymbol{A})^{-1} \boldsymbol{B}.
		\end{aligned}
	\end{equation*}
\end{definition}
If the matrices $\boldsymbol{V} \!$ and $\boldsymbol{W} \!$ are chosen as bases of res\-pective $r$-order \emph{input} and \emph{output rational Krylov subspaces}
\begin{subequations} \label{eq:multimom-block-Krylov}
	\begin{align}
		\mathcal{K}_r\left((\sigma \boldsymbol{E} \!-\! \boldsymbol{A})^{-1} \boldsymbol{E}, (\sigma \boldsymbol{E} \!-\! \boldsymbol{A})^{-1} \boldsymbol{B}\right) \subseteq \mathrm{Ran}(\boldsymbol{V}), \label{eq:multimom-block-Krylov-V} \\[0.2em] 
		\mathcal{K}_r\left((\mu \boldsymbol{E} \!-\! \boldsymbol{A})^{- \mathsf{T}} \boldsymbol{E}^{\mathsf T}, (\mu \boldsymbol{E} \!-\! \boldsymbol{A})^{- \mathsf{T}} \boldsymbol{C}^{\mathsf T}\right) \subseteq \mathrm{Ran}(\boldsymbol{W}),
	\end{align}
\end{subequations}
then the ROM \eqref{eq:linear-ROM} matches $r$ moments of the original transfer function around the expansion points $\sigma$ and $\mu$, respectively. 

In addition to the afore explained \emph{multimoment} matching strategy, note that it is also possible to match (high-order) moments at a set of different shifts $\left\{\sigma_i\right\}_{i=1}^q$ and $\left\{\mu_i\right\}_{i=1}^q$ with associated multiplicities $\left\{r_i\right\}_{i=1}^q$, where $\sum_{i=1}^{q} r_i = r$. In this setting, known as \emph{multipoint} moment matching, each subspace $\mathrm{Ran}(\boldsymbol{V})$ and $\mathrm{Ran}(\boldsymbol{W})$ is given by the union of all respective rational Krylov subspaces $\mathcal{K}_{r_i}$.
% This case is known as \emph{multipoint} moment matching.

Note also that, besides the \emph{block} Krylov subspaces \eqref{eq:multimom-block-Krylov}, in the MIMO case we alternatively may use so-called \emph{tangential} Krylov subspaces (e.g. for $r_1=\ldots=r_q = 1$):
\begin{subequations} \label{eq:multipoint-tang-Krylov}
	\begin{equation}
		\resizebox{0.89\linewidth}{!}{
			$\mathrm{span}\left\{
			(\sigma_1 \boldsymbol{E} \!-\! \boldsymbol{A})^{-1} \! \boldsymbol{B} \, \boldsymbol{r}_1, \ldots, (\sigma_r \boldsymbol{E} \!-\! \boldsymbol{A})^{-1} \! \boldsymbol{B} \, \boldsymbol{r}_r\right\} \subseteq \mathrm{Ran}(\boldsymbol{V})$,}
	\end{equation}
	%	\vspace{-2em}
	\begin{equation}
		\resizebox{0.89\linewidth}{!}{
			$\mathrm{span}\left\{
			(\mu_1 \boldsymbol{E} \!-\! \boldsymbol{A})^{- \! \mathsf T}\boldsymbol{C}^{\mathsf T} \boldsymbol{l}_1, \ldots, (\mu_r \boldsymbol{E} \!-\! \boldsymbol{A})^{- \! \mathsf T}\boldsymbol{C}^{\mathsf T} \boldsymbol{l}_r\right\} \subseteq \mathrm{Ran}(\boldsymbol{W})$.}
	\end{equation}
\end{subequations}
Here, convenient right and left tangential directions $\! \boldsymbol{r}_i \!\in\! \mathbb{C}^m$ and $\boldsymbol{l}_i \in \mathbb{C}^p$ are chosen to tangentially interpolate the transfer function at selected expansion points $\sigma_i$, $\mu_i \in \mathbb{C} \setminus \uplambda(\boldsymbol{E}^{-1} \boldsymbol{A})$.

%\begin{remark}(Two-sided, one-sided)
%	The choice 
%\end{remark}

\subsection{Equivalence of Krylov subspaces and Sylvester equations}
In fact, any basis of an input and output Krylov subspace \eqref{eq:multipoint-tang-Krylov} can be equivalently interpreted as the solution $\boldsymbol{V}$ and $\boldsymbol{W}$ of the following Sylvester equations \cite{gallivan2004sylvester}:
\begin{subequations}
	\begin{align}
		\boldsymbol{E} \, \boldsymbol{V} \, \boldsymbol{S}_{v} - \boldsymbol{A} \, \boldsymbol{V} &= \boldsymbol{B} \, \boldsymbol{R}\,, \label{eq:Sylv-V}\\
		\boldsymbol{E}^{\mathsf{T}} \, \boldsymbol{W} \, \boldsymbol{S}_{w}^{\mathsf T} - \boldsymbol{A}^{\mathsf{T}} \, \boldsymbol{W} &= \boldsymbol{C}^{\mathsf{T}} \, \boldsymbol{L}.
	\end{align}
\end{subequations}
Hereby, the input interpolation data $\left\{\sigma_i, \boldsymbol{r}_i\right\}$ is specified by the matrices $\boldsymbol{S}_v \!=\! \mathrm{diag}(\sigma_1, \ldots, \sigma_r) \!\in\! \mathbb{C}^{r \times r}$ and $\boldsymbol{R} \!=\! \left[\boldsymbol{r}_1, \ldots, \boldsymbol{r}_r\right] \in \mathbb{C}^{m \times r}$, where the pair $(\boldsymbol{R}, \boldsymbol{S}_v)$ is observable. Similarly, the output interpolation data $\left\{\mu_i, \boldsymbol{l}_i\right\}$ is denoted by the matrices $\boldsymbol{S}_w \!=\! \mathrm{diag}(\mu_1, \ldots, \mu_r) \in \mathbb{C}^{r \times r}$ and $\boldsymbol{L} \!=\! \left[\boldsymbol{l}_1, \ldots, \boldsymbol{l}_r \right] \in \mathbb{C}^{p \times r}$, where the pair $(\boldsymbol{S}_w, \boldsymbol{L}^{\mathsf T})$ is controllable.

Note that in the multimoment case, $\boldsymbol{S}_v, \boldsymbol{S}_w$ are Jordan matrices, and that in the SISO case\footnote{For SISO ($m\!=\!1, p\!=\!1$) replace $\boldsymbol{B} \to \boldsymbol{b} \in \mathbb{R}^{n}$ and $\boldsymbol{C} \to \boldsymbol{c}^{\mathsf T} \in \mathbb{R}^{1 \times n}$.}, $\boldsymbol{R}, \boldsymbol{L}$ become row vectors with corresponding ones and zeros. 

\subsection{Time domain interpretation of Moment Matching}
%Time-domain interpretation with interconnected systems and signal generator (input is an exponential function).
%Moments are the steady-state responses ...
In addition to the frequency domain perception of moment matching by means of the interpolation of the original transfer function around certain shifts, one can also interpret this concept in the time domain. % using the Sylvester equivalence.   
To this end, moments will be first characterized in terms of the solution of a Sylvester equation in Lemma \ref{th:lin-moments-Sylvester}, and then interpreted as the steady-state response of an interconnected system in Theorem \ref{th:lin-moments-steady-state} \cite{astolfi2010model,astolfi2010steady}. 
\begin{lemma} \label{th:lin-moments-Sylvester}
	The moments $\boldsymbol{m}_i(\sigma)$ of system \eqref{eq:linear-FOM} around shifts $\sigma \not\in \uplambda(\boldsymbol{E}^{-1} \boldsymbol{A})$ are equivalently given by
	\begin{equation}
		\boldsymbol{m}_i(\sigma) = (-1)^i \, \boldsymbol{C} \boldsymbol{V}_i\,, \quad i=0,\ldots,r-1
	\end{equation} \vspace{-1em}
	where, according to \eqref{eq:multimom-block-Krylov-V}, $\boldsymbol{V}_i$ is calculated as
	\begin{equation}
		\begin{aligned}
			(\sigma \boldsymbol{E} - \boldsymbol{A}) \boldsymbol{V}_0 &= \boldsymbol{B}, \\ %\, \boldsymbol{\mathrm{I}}_m, \\
			(\sigma \boldsymbol{E} - \boldsymbol{A}) \boldsymbol{V}_i &= \boldsymbol{E} \, \boldsymbol{V}_{i-1}, \quad i \geq 1 
		\end{aligned}
	\end{equation}
	or, alternatively, $\boldsymbol{V} \!=\! \left[\boldsymbol{V}_0, \ldots, \boldsymbol{V}_{r-1}\right]$ corresponds to the unique solution of the Sylvester equation \eqref{eq:Sylv-V}
	%	\begin{equation}
	%	\boldsymbol{E} \, \boldsymbol{V} \, \boldsymbol{S}_{v} - \boldsymbol{A} \, \boldsymbol{V} = \boldsymbol{B} \, \boldsymbol{R}
	%	\end{equation}
	with the corresponding ``modified" Jordan matrix $\boldsymbol{S}_v$ with \emph{negative} off-diagonal square blocks and $\boldsymbol{R} \!=\! \begin{bmatrix}
	\boldsymbol{\mathrm{I}}_m & \boldsymbol{0}_m & \cdots & \boldsymbol{0}_m
	\end{bmatrix}$.
\end{lemma}
\begin{theorem} \cite{astolfi2010model} \label{th:lin-moments-steady-state}
	Consider the interconnection of system \eqref{eq:linear-FOM} with the signal generator
	\begin{equation} \label{eq:lin-SG}
		\begin{aligned}
			\dot{\boldsymbol{x}}_{\mathrm{r}}^v(t) &= \boldsymbol{S}_{v} \, \boldsymbol{x}_{\mathrm{r}}^v(t), \quad \boldsymbol{x}_{\mathrm{r}}^v(0) = \boldsymbol{x}_{\mathrm{r},0}^v \neq \boldsymbol{0}, \\
			\boldsymbol{u}(t) &= \boldsymbol{R} \, \boldsymbol{x}_{\mathrm{r}}^v(t),
		\end{aligned}
	\end{equation}
	where the triple $(\boldsymbol{S}_v, \, \boldsymbol{R}, \, \boldsymbol{x}_{\mathrm{r},0}^v)$ is such that $(\boldsymbol{R}, \boldsymbol{S}_v)$ is observable, $\uplambda(\boldsymbol{S}_v) \, \cap \, \uplambda(\boldsymbol{E}^{-1} \boldsymbol{A}) \!=\! \emptyset$ and $(\boldsymbol{S}_v, \boldsymbol{x}_{\mathrm{r},0}^v)$ is excitable. Then, the moments of system \eqref{eq:linear-FOM} are related to the (well-defined) steady-state response of the output $\boldsymbol{y}(t) \!=\! \boldsymbol{y}_{\mathrm{r}}(t) \!=\! \boldsymbol{C} \boldsymbol{V} \boldsymbol{x}_{\mathrm{r}}^v(t)$ of such interconnected system (cf. Fig~\ref{fig:lin-sys-lin-SG_steady-state-V}), where $\boldsymbol{V}$ is the unique solution of the Sylvester equation \eqref{eq:Sylv-V}. 
\end{theorem}
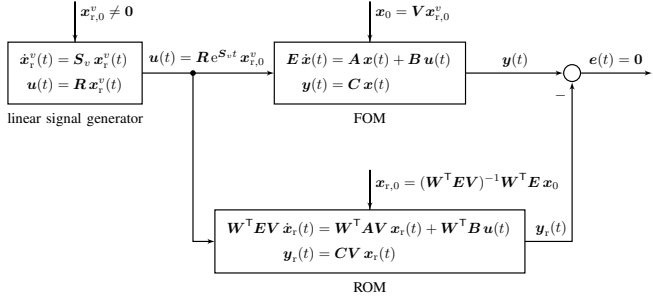
\begin{figure}[tp]
	\centering
	\scalebox{0.56}{
			\input{tikz/lin-sys-lin-SG_steady-state-V}}
	\caption{\footnotesize Diagram depicting the interconnection between the linear FOM/ROM and the linear signal generator to illustrate the time domain interpretation of moment matching for linear systems.}
	\label{fig:lin-sys-lin-SG_steady-state-V}
\end{figure} %\vspace{-1em}
\begin{corollary}
	Interconnecting system \eqref{eq:linear-FOM} with the signal ge\-nerator \eqref{eq:lin-SG} is equivalent to exciting the FOM with exponential input signals $\boldsymbol{u}(t) \!=\! \boldsymbol{R} \, \boldsymbol{x}_{\mathrm{r}}^v(t) \!=\! \boldsymbol{R} \, \mathrm{e}^{\boldsymbol{S}_v t} \, \boldsymbol{x}_{\mathrm{r},0}^v$ with exponents given by the shift matrix $\boldsymbol{S}_v$. Consequently, given $\boldsymbol{u}(t) \!=\! \boldsymbol{R} \, \mathrm{e}^{\boldsymbol{S}_v t} \, \boldsymbol{x}_{\mathrm{r},0}^v$ with $\boldsymbol{x}_{0} \!=\! \boldsymbol{V} \boldsymbol{x}_{\mathrm{r},0}^v$, $\boldsymbol{x}_{\mathrm{r},0}^v \!\neq\! \boldsymbol{0}$ arbitrary, $\boldsymbol{V}$ as solution of \eqref{eq:Sylv-V} and $\boldsymbol{W}$ such that $\det(\boldsymbol{W}^{\mathsf T} \boldsymbol{E} \boldsymbol{V}) \neq 0$, then the (asymptotically stable) ROM \eqref{eq:linear-ROM} exactly matches the steady-state response of the output of the FOM, i.e. $\boldsymbol{e}(t) \!=\! \boldsymbol{y}(t) - \boldsymbol{y}_{\mathrm{r}}(t) \!=\! \boldsymbol{C} \boldsymbol{x}(t) - \boldsymbol{C} \boldsymbol{V} \boldsymbol{x}_{\mathrm{r}}(t)\!=\! \boldsymbol{0} \ \forall \, t$ (see Fig. \ref{fig:lin-sys-lin-SG_steady-state-V}).
\end{corollary}

Thus, linear moment matching can be interpreted as the interpolation of the steady-state response of the output of the FOM, when this is excited with exponential input signals. This understanding follows from the fact that the transfer function $G(s)$ represents the scaling factor of (complex) exponentials $\mathrm{e}^{st}$, which are the \emph{eigenfunctions} of LTI systems, i.e. $y(t) \!=\! G(s) \, \mathrm{e}^{s t}$ for $u(t) \!=\! \mathrm{e}^{st}$. Interestingly enough, the Sylvester equation \eqref{eq:Sylv-V} can be alternatively derived using this time domain perception and the notion of signal generators. To this end, first insert the linear approximation ansatz $\boldsymbol{x}(t) \!=\! \boldsymbol{V} \boldsymbol{x}_{\mathrm{r}}(t)$ with $\boldsymbol{x}_{\mathrm{r}}(t) \!\overset{!}{=}\! \boldsymbol{x}_{\mathrm{r}}^v(t)$ in the state equation of~\eqref{eq:linear-FOM}:
\begin{equation} \label{eq:derivation-Syl}
	\boldsymbol{E} \, \boldsymbol{V} \, \dot{\boldsymbol{x}}_{\mathrm{r}}^v(t) = \boldsymbol{A} \, \boldsymbol{V} \, \boldsymbol{x}_{\mathrm{r}}^v(t) + \boldsymbol{B} \, \boldsymbol{u}(t).
\end{equation} 
Subsequently, the linear signal generator $\dot{\boldsymbol{x}}_{\mathrm{r}}^v(t) \!=\! \boldsymbol{S}_{v} \, \boldsymbol{x}_{\mathrm{r}}^v(t)$, $\boldsymbol{u}(t) \!=\! \boldsymbol{R} \, \boldsymbol{x}_{\mathrm{r}}^v(t)$ is plugged into \eqref{eq:derivation-Syl}, yielding
\begin{equation} \label{eq:derivation-Syl-2}
	\boldsymbol{0} = \left(\boldsymbol{A} \, \boldsymbol{V} - \boldsymbol{E} \, \boldsymbol{V} \, \boldsymbol{S}_{v} + \boldsymbol{B} \, \boldsymbol{R}\right) \cdot \boldsymbol{x}_{\mathrm{r}}^v(t).
\end{equation}
%Since this equation holds for all $\boldsymbol{x}_{\mathrm{r}}(t)\neq \boldsymbol{0}$, the state vector $\boldsymbol{x}_{\mathrm{r}}(t)$ can be factored out and a \emph{constant} (state-independent) linear Sylvester equation of dimension $n \times r$ is obtained
Since the equation holds for $\boldsymbol{x}_{\mathrm{r}}^v(t) \!\!=\! \mathrm{e}^{\boldsymbol{S}_v t} \boldsymbol{x}_{\mathrm{r},0}^v$, the state vector $\boldsymbol{x}_{\mathrm{r}}^v(t)$ can be factored out and a \emph{constant} (state-independent) linear Sylvester equation of dimension $n \!\times\! r$ is obtained
\begin{equation}
	\boldsymbol{A} \, \boldsymbol{V} - \boldsymbol{E} \, \boldsymbol{V} \, \boldsymbol{S}_{v} + \boldsymbol{B} \, \boldsymbol{R} = \boldsymbol{0},
\end{equation}
whose solution $\boldsymbol{V}$ spans a corresponding input rational Krylov subspace which guarantees moment matching under the aforementioned circumstances. 

%%%%%%%%%%%%%%%%%%%%%%%%%%%%%%%%%%%%%%%%%%%%%%%%%%%%%%%%%%%%%%%%%%%%%%%%%%%%%%%%
\section{Moment Matching for nonlinear systems} \label{sec:nlmm-PDE}

\subsection{Nonlinear Petrov-Galerkin projection}
Consider now a large-scale, nonlinear time-invariant, exponentially stable, MIMO state-space model of the form
\begin{equation} \label{eq:nonlin-FOM}
	\begin{aligned}
		\dot{\boldsymbol{x}}(t) &= \boldsymbol{f}\big(\boldsymbol{x}(t), \boldsymbol{u}(t)\big), \quad \boldsymbol{x}(0) = \boldsymbol{x}_0,\\
		\boldsymbol{y}(t) &= \boldsymbol{h}\big(\boldsymbol{x}(t)\big),
	\end{aligned}
\end{equation}
with $\boldsymbol{x}(t) \in \mathbb{R}^n$, $\boldsymbol{u}(t) \in \mathbb{R}^m$, $\boldsymbol{y}(t) \in \mathbb{R}^p$ and smooth mappings $\boldsymbol{f}(\boldsymbol{x}, \boldsymbol{u}): \mathbb{R}^n \times \mathbb{R}^m \to \mathbb{R}^n$ and $\boldsymbol{h}(\boldsymbol{x}) : \mathbb{R}^n \to \mathbb{R}^p$, such that $\boldsymbol{f}(\boldsymbol{0}, \boldsymbol{0}) \!=\! \boldsymbol{0}$ and $\boldsymbol{h}(\boldsymbol{0}) \!=\! \boldsymbol{0}$. The aim now is to find a nonlinear ROM of dimension $r \ll n$ using again a projection framework. %within a projection framework
One established and successful way to do this, is by applying the classical Petrov-Galerkin projection with linear mappings given by the matrices $\boldsymbol{V}, \boldsymbol{W}$ to the nonlinear system \eqref{eq:nonlin-FOM}. Herein, the projection matrices are generally constructed via POD or other nonlinear reduction techniques. Another possibility is to apply a nonlinear Petrov-Galerkin projection using \emph{nonlinear mappings} defined on \emph{manifolds} \cite{ionescu2016nonlinear}. To this end, let
$\boldsymbol{x}(t) \!=\! \boldsymbol{\nu}(\boldsymbol{x}_{\mathrm{r}}(t))$ be the nonlinear approximation ansatz with smooth mapping $\boldsymbol{\nu}(\boldsymbol{x}_{\mathrm{r}}) : \mathbb{R}^r \to \mathbb{R}^n$. Furthermore, define $\boldsymbol{x}_{\mathrm{r}}(t) \!=\! \boldsymbol{\omega}(\boldsymbol{x}(t))$ with smooth mapping $\boldsymbol{\omega}(\boldsymbol{x}) : \mathbb{R}^n \to \mathbb{R}^r$, such that $\boldsymbol{\omega}(\boldsymbol{\nu}(\boldsymbol{x}_{\mathrm{r}})) \!=\! \boldsymbol{x}_{\mathrm{r}}$. The reduction is then performed through a nonlinear Petrov-Galerkin projection $\boldsymbol{\varrho}(\boldsymbol{x}(t)) \!=\! \boldsymbol{\nu}\big(\boldsymbol{\omega}(\boldsymbol{x}(t))\big)$ of \eqref{eq:nonlin-FOM} by means of the projector mapping $\boldsymbol{\varrho}(\boldsymbol{x}) : \mathbb{R}^n \to \mathbb{R}^n$, yielding the nonlinear ROM
\begin{equation} \label{eq:nonlin-ROM}
	\begin{aligned}
		\dot{\boldsymbol{x}}_{\mathrm{r}}(t) &= \left.\frac{\partial \boldsymbol{\omega}(\boldsymbol{x}(t))}{\partial \boldsymbol{x}(t)} \, \boldsymbol{f}\big(\boldsymbol{x}(t),  \boldsymbol{u}(t)\big)\right|_{\boldsymbol{x}(t)=\boldsymbol{\nu}(\boldsymbol{x}_{\mathrm{r}}(t))}\,, \\
		\boldsymbol{y}_{\mathrm{r}}(t) &= \boldsymbol{h}\big(\boldsymbol{\nu}(\boldsymbol{x}_{\mathrm{r}}(t))\big),
	\end{aligned}
\end{equation}
where $\boldsymbol{x}_{\mathrm{r}}(t) \! \in \! \mathbb{R}^r$, $(\partial \boldsymbol{\omega}(\boldsymbol{x})/\partial \boldsymbol{x})|_{\boldsymbol{x}\!=\!\boldsymbol{\nu}(\boldsymbol{x}_{\mathrm{r}})} \!\cdot\! (\partial \boldsymbol{\nu}(\boldsymbol{x}_{\mathrm{r}})/\partial \boldsymbol{x}_{\mathrm{r}}) \!=\! \mathbf{I}_r$ and the initial condition is $\boldsymbol{x}_{\mathrm{r}}(0) \!=\! \boldsymbol{\omega}(\boldsymbol{x}_0)$.
\begin{remark}
	The afore explained nonlinear projection framework (for $\boldsymbol{E} \!=\! \mathbf{I}$) is a generalization from the linear case. Therefore, the nonlinear mappings are strongly related to their linear counterparts:
	\begin{subequations}
		\vspace{-0.6em}
		\begin{align}
			\boldsymbol{x} &= \boldsymbol{\nu}(\boldsymbol{x}_{\mathrm{r}}) \ \widehat{=} \ \boldsymbol{V} \, \boldsymbol{x}_{\mathrm{r}}, && \frac{\partial \boldsymbol{\nu}(\boldsymbol{x}_{\mathrm{r}})}{\partial \boldsymbol{x}_{\mathrm{r}}} \ \widehat{=} \ \boldsymbol{V},\\[0.1em]
			\boldsymbol{x}_{\mathrm{r}} &= \boldsymbol{\omega}(\boldsymbol{x}) \ \widehat{=} \ \underbrace{(\boldsymbol{W}^{\mathsf T}\boldsymbol{V})^{-1}\boldsymbol{W}^{\mathsf T}}_{*} \boldsymbol{x}, && \ \frac{\partial \boldsymbol{\omega}(\boldsymbol{x})}{\partial \boldsymbol{x}} \ \widehat{=} \ * \,, 
			\\[-0.3em]
			\boldsymbol{\varrho} &= \boldsymbol{\nu}\big(\boldsymbol{\omega}(\boldsymbol{x})\big) \ \widehat{=} \ \boldsymbol{P} \boldsymbol{x}, && \!\!\!\!\!\!\!\!\!\!\!\!\!\!\!\!\! \frac{\partial \boldsymbol{\nu}(\boldsymbol{x}_{\mathrm{r}})}{\partial \boldsymbol{x}_{\mathrm{r}}} \frac{\partial \boldsymbol{\omega}(\boldsymbol{x})}{\partial \boldsymbol{x}} \ \widehat{=} \ \boldsymbol{P}.
		\end{align}
	\end{subequations}
%	Note that the condition $\boldsymbol{\omega}(\boldsymbol{\nu}(\boldsymbol{x}_{\mathrm{r}})) \!\!=\!\! \boldsymbol{x}_{\mathrm{r}}$ corresponds to $(\boldsymbol{W}^{\mathsf T}\boldsymbol{V})^{-1}\boldsymbol{W}^{\mathsf T} \boldsymbol{V} \boldsymbol{x}_{\mathrm{r}} \!\!=\!\! \boldsymbol{x}_{\mathrm{r}}$. More\-over, for the nonlinear projection mapping it holds $\boldsymbol{\varrho}(\boldsymbol{x}) \, \widehat{=} \, \boldsymbol{P} \, \boldsymbol{x}$, which corresponds to the linear ansatz with the linear projector $\boldsymbol{P} \!=\! \boldsymbol{V}(\boldsymbol{W}^{\mathsf T}\boldsymbol{V})^{-1}\boldsymbol{W}^{\mathsf T}$. For nonlinear systems, a one-sided projection ($\boldsymbol{W} \!=\! \boldsymbol{V}$) is commonly used, which yields $*\!=\!\boldsymbol{V}^{\mathsf T}$, $\boldsymbol{P} \!=\! \boldsymbol{V}\boldsymbol{V}^{\mathsf T}$ and $\boldsymbol{x}_{\mathrm{r},0} \!=\! \boldsymbol{V}^{\mathsf T}\boldsymbol{x}_0$, provided that $\boldsymbol{V}$ is orthogonal ($\boldsymbol{V}^{\mathsf T} \boldsymbol{V} \!=\! \boldsymbol{\mathrm{I}}_r$).	
	Note that the condition $\boldsymbol{\omega}(\boldsymbol{\nu}(\boldsymbol{x}_{\mathrm{r}})) \!\!=\!\! \boldsymbol{x}_{\mathrm{r}}$ corresponds to $(\boldsymbol{W}^{\mathsf T}\boldsymbol{V})^{-1}\boldsymbol{W}^{\mathsf T} \boldsymbol{V} \boldsymbol{x}_{\mathrm{r}} \!\!=\!\! \boldsymbol{x}_{\mathrm{r}}$ and that $\boldsymbol{P} \!=\! \boldsymbol{V}(\boldsymbol{W}^{\mathsf T}\boldsymbol{V})^{-1}\boldsymbol{W}^{\mathsf T}$. For nonlinear systems, a one-sided projection ($\boldsymbol{W} \!=\! \boldsymbol{V}$) is commonly used, which yields $*\!=\!\boldsymbol{V}^{\mathsf T}$, $\boldsymbol{P} \!=\! \boldsymbol{V}\boldsymbol{V}^{\mathsf T}$ and $\boldsymbol{x}_{\mathrm{r},0} \!=\! \boldsymbol{V}^{\mathsf T}\boldsymbol{x}_0$, provided that $\boldsymbol{V}$ is orthogonal ($\boldsymbol{V}^{\mathsf T} \boldsymbol{V} \!=\! \boldsymbol{\mathrm{I}}_r$).	
\end{remark}  
\begin{figure}[tp]
	\centering
	\scalebox{0.56}{
			\input{tikz/nonlin-sys-nonlin-SG_steady-state-V}}
	\caption{\footnotesize Diagram depicting the interconnection between the nonlinear FOM/ROM and the nonlinear signal generator to illustrate the time domain interpretation of moment matching for nonlinear systems.}
	\label{fig:nonlin-sys-nonlin-SG_steady-state-V}
\end{figure}
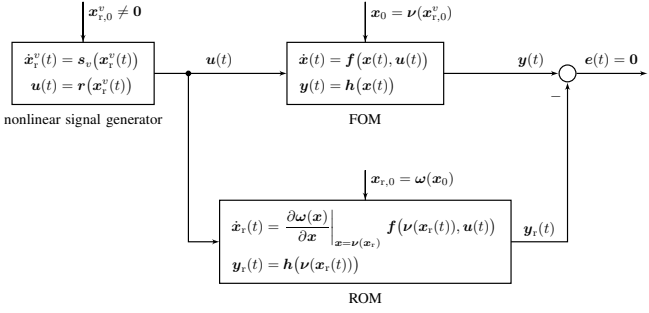
\subsection{Notion of Nonlinear Moments}
The notion of moments and their steady-state-based interpretation can be carried over to nonlinear systems \cite{astolfi2010model}. %To achieve this, consider the following nonlinear signal generator
\begin{theorem} \cite{astolfi2010model} \label{th:nonlin-moments-steady-state}
	Consider the interconnection of system \eqref{eq:nonlin-FOM} with the nonlinear signal generator
	\begin{equation} \label{eq:nonlin-SG}
		\begin{aligned}
			\dot{\boldsymbol{x}}_{\mathrm{r}}^v(t) &= \boldsymbol{s}_{v}\big(\boldsymbol{x}_{\mathrm{r}}^v(t)\big), \quad \boldsymbol{x}_{\mathrm{r}}^v(0) = \boldsymbol{x}_{\mathrm{r},0}^v \neq \boldsymbol{0}, \\
			\boldsymbol{u}(t) &= \boldsymbol{r}\big(\boldsymbol{x}_{\mathrm{r}}^v(t)\big),
		\end{aligned}
	\end{equation}
	where $\boldsymbol{s}_{v}(\boldsymbol{x}_{\mathrm{r}}^v) \!:\! \mathbb{R}^r \to \mathbb{R}^r$, $\boldsymbol{r}(\boldsymbol{x}_{\mathrm{r}}^v) \!:\! \mathbb{R}^r \to \mathbb{R}^m$ are smooth mappings such that $\boldsymbol{s}_v(\boldsymbol{0}) \!=\! \boldsymbol{0}$ and $\boldsymbol{r}(\boldsymbol{0}) \!=\! \boldsymbol{0}$. Hereby it is assumed, that the signal generator $(\boldsymbol{r}, \, \boldsymbol{s}_v, \, \boldsymbol{x}_{\mathrm{r},0}^v)$ is ob\-ser\-va\-ble, i.e. for any pair of initial conditions $\boldsymbol{x}_{\mathrm{r},\mathrm{a}}^v(0) \! \neq \! \boldsymbol{x}_{\mathrm{r},\mathrm{b}}^v(0)$, the corresponding trajectories $\boldsymbol{r}(\boldsymbol{x}_{\mathrm{r},\mathrm{a}}^v(t))$ and $\boldsymbol{r}(\boldsymbol{x}_{\mathrm{r},\mathrm{b}}^v(t))$ do not coincide. In addition, the signal generator \eqref{eq:nonlin-SG} can be Poisson stable in a neighborhood of its equilibrium $\boldsymbol{x}_{\mathrm{r}}^v \!=\! \boldsymbol{0}$ with $\boldsymbol{x}_{\mathrm{r}}^v(0) \! \neq \! \boldsymbol{0}$. Further assume that the zero equilibrium of the system $\dot{\boldsymbol{x}} \!=\! \boldsymbol{f}(\boldsymbol{x}, \boldsymbol{0})$ is locally exponentially stable. Then, the moments of system \eqref{eq:nonlin-FOM} at $(\boldsymbol{s}_v(\boldsymbol{x}_{\mathrm{r}}^v), \boldsymbol{r}(\boldsymbol{x}_{\mathrm{r}}^v), \boldsymbol{x}_{\mathrm{r},0}^v)$ are related to the (locally well-defined) steady-state response of the output $\boldsymbol{y}(t) \!=\! \boldsymbol{y}_{\mathrm{r}}(t) \!=\! \boldsymbol{h}\big(\boldsymbol{\nu}(\boldsymbol{x}_{\mathrm{r}}^v(t))\big)$ of such in\-ter\-co\-nnec\-ted system (cf. Fig. \ref{fig:nonlin-sys-nonlin-SG_steady-state-V}), where the mapping $\boldsymbol{\nu}(\boldsymbol{x}_{\mathrm{r}}^v)$, defined in a neighborhood of $\boldsymbol{x}_{\mathrm{r}}^v \!=\! \boldsymbol{0}$, is the unique solution of the following Sylvester-like partial differential equation~(PDE)
	\begin{equation} \label{eq:nonlin-Sylv-v}
		\frac{\partial \boldsymbol{\nu}(\boldsymbol{x}_{\mathrm{r}}^v)}{\partial \boldsymbol{x}_{\mathrm{r}}^v} \, \boldsymbol{s}_v(\boldsymbol{x}_{\mathrm{r}}^v) = \boldsymbol{f}\big(\boldsymbol{\nu}(\boldsymbol{x}_{\mathrm{r}}^v), \boldsymbol{r}(\boldsymbol{x}_{\mathrm{r}}^v)\big).
	\end{equation}
\end{theorem}
%TODO: remark with importance of signal generator!?
\vspace{0.2em}
\subsection{Steady-State-Based Nonlinear Moment Matching} %Steady-State-Based perception of moment matching
Based on Theorem \ref{th:nonlin-moments-steady-state}, the perception of nonlinear moment matching in terms of the interpolation of the steady-state response of an interconnected system follows.
\begin{corollary}
	Consider the interconnection of system \eqref{eq:nonlin-FOM} and the nonlinear signal generator \eqref{eq:nonlin-SG}. Suppose all assumptions concerning observability and local exponential stability from above hold. Moreover, let $\boldsymbol{\nu}(\boldsymbol{x}_{\mathrm{r}}^v)$ be the unique solution of \eqref{eq:nonlin-Sylv-v} and $\boldsymbol{\omega}(\boldsymbol{x})$ such that $\boldsymbol{\omega}(\boldsymbol{\nu}(\boldsymbol{x}_{\mathrm{r}}^v)) \!=\! \boldsymbol{x}_{\mathrm{r}}^v$. Assume that the zero equilibrium of \eqref{eq:nonlin-ROM} is locally exponentially stable. Then, the ROM \eqref{eq:nonlin-ROM} exactly matches the steady-state response of the output of the FOM, i.e. $\boldsymbol{e}(t) \!=\! \boldsymbol{y}(t) - \boldsymbol{y}_{\mathrm{r}}(t) \!=\! \boldsymbol{h}\big(\boldsymbol{x}(t)\big) - \boldsymbol{h}\big(\boldsymbol{\nu}(\boldsymbol{x}_{\mathrm{r}}(t))\big) \!=\! \boldsymbol{0} \ \forall \, t$ (see Fig. \ref{fig:nonlin-sys-nonlin-SG_steady-state-V}).
\end{corollary}

Note that the Sylvester-like PDE from \eqref{eq:nonlin-Sylv-v} represents the nonlinear counterpart of the linear equation \eqref{eq:derivation-Syl-2}:
\begin{equation}
	\boldsymbol{V} \, \boldsymbol{S}_v \, \mathrm{e}^{\boldsymbol{S}_v t} \, \boldsymbol{x}_{\mathrm{r},0}^v = \boldsymbol{A} \, \boldsymbol{V} \, \mathrm{e}^{\boldsymbol{S}_v t} \, \boldsymbol{x}_{\mathrm{r},0}^v + \boldsymbol{B} \, \boldsymbol{R} \, \mathrm{e}^{\boldsymbol{S}_v t} \, \boldsymbol{x}_{\mathrm{r},0}^v.
\end{equation}
Thus, the PDE can be alternatively derived as follows. First, the nonlinear approximation ansatz $\boldsymbol{x}(t) \!=\! \boldsymbol{\nu}(\boldsymbol{x}_{\mathrm{r}}(t))$ with $\boldsymbol{x}_{\mathrm{r}}(t) \overset{!}{=} \boldsymbol{x}_{\mathrm{r}}^v(t)$ is inserted in the state equation of \eqref{eq:nonlin-FOM}: 
\begin{equation} \label{eq:derivation-PDE}
	\frac{\partial \boldsymbol{\nu}(\boldsymbol{x}_{\mathrm{r}}^v(t))}{\partial \boldsymbol{x}_{\mathrm{r}}^v(t)} \, \dot{\boldsymbol{x}}_{\mathrm{r}}^v(t) = \boldsymbol{f}\big(\boldsymbol{\nu}(\boldsymbol{x}_{\mathrm{r}}^v(t)), \boldsymbol{u}(t)\big).
\end{equation}
Afterwards, the nonlinear signal generator $\dot{\boldsymbol{x}}_{\mathrm{r}}^v(t) \!\!=\!\! \boldsymbol{s}_{v}\big(\boldsymbol{x}_{\mathrm{r}}^v(t)\big)$, $\boldsymbol{u}(t) \!=\! \boldsymbol{r}\big(\boldsymbol{x}_{\mathrm{r}}^v(t)\big)$ is plugged into \eqref{eq:derivation-PDE}, yielding
\begin{equation} \label{eq:nonlin-time-dep-Sylv-v}
	\frac{\partial \boldsymbol{\nu}(\boldsymbol{x}_{\mathrm{r}}^v(t))}{\partial \boldsymbol{x}_{\mathrm{r}}^v(t)} \, \boldsymbol{s}_{v}\big(\boldsymbol{x}_{\mathrm{r}}^v(t)\big) = \boldsymbol{f}\big(\boldsymbol{\nu}(\boldsymbol{x}_{\mathrm{r}}^v(t)), \boldsymbol{r}(\boldsymbol{x}_{\mathrm{r}}^v(t))\big).
\end{equation}  
Note that -- in contrast to the linear, state-independent Sylvester equation \eqref{eq:Sylv-V} of dimension $n \times r$ -- the PDE \eqref{eq:nonlin-time-dep-Sylv-v} is a \emph{nonlinear}, \emph{state-dependent} equation of dimension $n \times 1$.

%%%%%%%%%%%%%%%%%%%%%%%%%%%%%%%%%%%%%%%%%%%%%%%%%%%%%%%%%%%%%%%%%%%%%%%%%%%%%%%%
\section{Simulation-Free Nonlinear Model Reduction by Moment Matching} \label{sec:simulation-free-NLMM}
The approach for nonlinear moment matching described in Section \ref{sec:nlmm-PDE} requires the solution $\boldsymbol{\nu}(\boldsymbol{x}_{\mathrm{r}}^v(t))$ of the nonlinear, state-dependent PDE \eqref{eq:nonlin-time-dep-Sylv-v} for a given signal generator, in order to reduce the FOM \eqref{eq:nonlin-FOM}. This involves either symbolic computations, or the numerical integration of a resulting system of ordinary differential equations (ODEs) after reduced state-space discretization of the PDE \eqref{eq:nonlin-time-dep-Sylv-v}. Since we aim to reduce large-scale nonlinear systems, almost only \emph{numerical} methods come into consideration, which preferably should also avoid an expensive simulation. Hence, some step-by-step simplifications are performed in the following towards a practicable, simulation-free method for nonlinear moment matching, which relies on the solution of a system of nonlinear \emph{algebraic} equations rather than of a PDE.  

% partially questionable assumptions 

% simplifications are performed step by step towards a practicable method for moment matching for nonlinear systems 

%equation above is not practicable since it involves the solution of a PDE, therefore -> use linear projection instead of nonlinear projection -> we get an algebraic equation 

\subsection{Linear Projection}
The first step towards a practical method consists in applying the linear projection ansatz $\boldsymbol{x}(t) \!=\! \boldsymbol{\nu}(\boldsymbol{x}_{\mathrm{r}}^v(t)) \!=\! \boldsymbol{V} \, \boldsymbol{x}_{\mathrm{r}}^v(t)$ instead of the nonlinear projection mapping $\boldsymbol{\nu}(\boldsymbol{x}_{\mathrm{r}}^v(t))$. This simplification is motivated by the fact that nonlinear projections are much more involved than linear ones, and that the latter are successfully employed even in nonlinear model order reduction. By doing so, the PDE \eqref{eq:nonlin-time-dep-Sylv-v} also becomes an algebraic equation, which is much easier to handle. Depending on the form of the used signal generator, we distinguish (based on \cite{astolfi2010model}) between the following cases:
\subsubsection{Nonlinear signal generator}
In this case, the PDE \eqref{eq:nonlin-time-dep-Sylv-v} becomes the following nonlinear system of equations
\begin{equation} \label{eq:LP-NSG}
	\boldsymbol{0} = \boldsymbol{f}\big(\boldsymbol{V} \boldsymbol{x}_{\mathrm{r}}^v(t), \boldsymbol{r}(\boldsymbol{x}_{\mathrm{r}}^v(t))\big) - \boldsymbol{V} \, \boldsymbol{s}_{v}\big(\boldsymbol{x}_{\mathrm{r}}^v(t)\big),
\end{equation}
where the triple $(\boldsymbol{s}_v(\boldsymbol{x}_{\mathrm{r}}^v(t)), \boldsymbol{r}(\boldsymbol{x}_{\mathrm{r}}^v(t)), \boldsymbol{x}_{\mathrm{r},0}^v)$ is user-defined and the projection matrix $\boldsymbol{V} \in \mathbb{R}^{n \times r}$ is the searched solution. Note, however, that system \eqref{eq:LP-NSG} consists of $n$ equations for $n \cdot r$ unknowns, i.e. it is underdetermined. Thus, we consider the equation column-wise for each $\boldsymbol{v}_i \in \mathbb{R}^n$, $i=1,\ldots,r$
\begin{equation} \label{eq:LP-NSG-elem}
	\boldsymbol{0} = \boldsymbol{f}\big(\boldsymbol{v}_i \, x_{\mathrm{r}, i}^v(t), \boldsymbol{r}_{_i}(x_{\mathrm{r},i}^v(t))\big) - \boldsymbol{v}_i \, s_{v_{i}}\big(x_{\mathrm{r},i}^v(t)\big),
\end{equation} 
with $x_{\mathrm{r},i}^v(t) \in \mathbb{R}$ and $\boldsymbol{V} \!=\! \left[\boldsymbol{v}_1, \ldots, \boldsymbol{v}_{r}\right]$. Please bear in mind that, in the linear setting, a column-wise construction of the orthogonal basis $\boldsymbol{V}$ using the Arnoldi process still fulfills the Sylvester matrix equation \eqref{eq:Sylv-V}. In the nonlinear setting, however, this does not hold true anymore, since equation \eqref{eq:LP-NSG} is generally not satisfied, even if each column $\boldsymbol{v}_i$ fulfills \eqref{eq:LP-NSG-elem}. This shortcoming is a consequence of the usage of a linear projection instead of a nonlinear mapping on a manifold.

\subsubsection{Linear signal generator} Motivated from the linear case, one may also come to the idea of interconnecting the nonlinear system \eqref{eq:nonlin-FOM} with the linear signal generator \eqref{eq:lin-SG}, where $\boldsymbol{s}_v(\boldsymbol{x}_{\mathrm{r}}^v(t)) \!=\! \boldsymbol{S}_v \, \boldsymbol{x}_{\mathrm{r}}^v(t)$ and $\boldsymbol{r}(\boldsymbol{x}_{\mathrm{r}}^v(t)) \!=\! \boldsymbol{R} \, \boldsymbol{x}_{\mathrm{r}}^v(t)$. 
By doing so, equation \eqref{eq:LP-NSG} becomes
\begin{equation}
	\boldsymbol{0} = \boldsymbol{f}\big(\boldsymbol{V} \boldsymbol{x}_{\mathrm{r}}^v(t), \boldsymbol{R} \, \boldsymbol{x}_{\mathrm{r}}^v(t)\big) - \boldsymbol{V} \, \boldsymbol{S}_{v} \, \boldsymbol{x}_{\mathrm{r}}^v(t),
\end{equation}
where the triple $(\boldsymbol{S}_v, \, \boldsymbol{R}, \, \boldsymbol{x}_{\mathrm{r},0}^v)$ is user-defined. Remember that the usage of a linear signal generator corresponds to exciting the nonlinear system with exponential input signals $\boldsymbol{u}(t) \!=\! \boldsymbol{R} \, \boldsymbol{x}_{\mathrm{r}}^v(t) \!=\! \boldsymbol{R} \, \mathrm{e}^{\boldsymbol{S}_v t} \, \boldsymbol{x}_{\mathrm{r},0}^v$. This choice naturally raises the question whether (growing) exponential inputs are sufficiently valid for characterizing nonlinear systems. Note that the dynamics of the selected signal generator represent the dynamics of the nonlinear system for which the steady-state responses are matched. Therefore, the signal generator should ideally be chosen such that it excites and cha\-rac\-te\-ri\-zes the important dynamics of the nonlinear system. It is well known that exponential functions are the characterizing eigenfunctions for linear systems. By exciting the nonlinear system with exponential input signals, we therefore hope to describe the nonlinear dynamics adequately as well.   

%Note that the dynamics of the chosen signal generator should ideally excite and represent the important dynamics of the nonlinear system, whose steady-state responses we want to match.  

Considering the underdetermined equation again column-wise delivers
\begin{equation} \label{eq:LP-LSG-elem}
	\boldsymbol{0} = \boldsymbol{f}\big( \boldsymbol{v}_i \, x_{\mathrm{r},i}^v(t), \underbrace{\boldsymbol{r}_i \, x_{\mathrm{r},i}^v(t)}_{\boldsymbol{r}_{_i}\left(x_{\mathrm{r},i}^v(t)\right)} \big) - \boldsymbol{v}_i \, \underbrace{\sigma_i \, x_{\mathrm{r},i}^v(t)}_{s_{v_{i}}\left(x_{\mathrm{r},i}^v(t)\right)},
\end{equation}
where the signal generator \eqref{eq:lin-SG} becomes $\dot{x}_{\mathrm{r},i}^v(t) \!=\! \sigma_i \, x_{\mathrm{r},i}^v(t)$, $\boldsymbol{u}_i(t) \!=\! \boldsymbol{r}_i \, x_{\mathrm{r},i}^v(t)$ with $x_{\mathrm{r},i}^v(t) \!=\! \mathrm{e}^{\sigma_i t} x_{\mathrm{r},0,i}^v$ for $i=1,\ldots,r$.

\subsubsection{Zero signal generator} This special (linear) signal generator is defined as $\dot{\boldsymbol{x}}_{\mathrm{r}}^v(t) \!=\! \boldsymbol{s}_v(\boldsymbol{x}_{\mathrm{r}}^v(t)) \!=\! \boldsymbol{0}$, which means that $\boldsymbol{x}_{\mathrm{r}}^v(t) \!=\! \boldsymbol{x}_{\mathrm{r},0}^v \!=\! \textrm{const}$ and $\boldsymbol{u}(t) \!=\! \boldsymbol{R} \boldsymbol{x}_{\mathrm{r}}^v(t) \!=\! \boldsymbol{R} \, \boldsymbol{x}_{\mathrm{r},0}^v \!=\! \textrm{const}$. Hence, the usage of a zero signal generator is equivalent to exciting the nonlinear system with a constant input signal.
In this particular case, equation \eqref{eq:LP-NSG} becomes
\begin{equation}
	\boldsymbol{0} = \boldsymbol{f}\big(\boldsymbol{V} \boldsymbol{x}_{\mathrm{r},0}^v, \ \boldsymbol{R} \, \boldsymbol{x}_{\mathrm{r},0}^v\big),
\end{equation}
which is a nonlinear, \emph{time-independent} system of equations.

A column-wise consideration of the underdetermined equation yields \vspace{-1.5em} 
\begin{equation} \label{eq:LP-ZSG-elem}
	\boldsymbol{0} = \boldsymbol{f}\big(\boldsymbol{v}_i \, x_{\mathrm{r},0,i}^v, \ \overbrace{\boldsymbol{r}_i \, x_{\mathrm{r},0,i}^v}^{\boldsymbol{r}_{_i}(x_{\mathrm{r},0,i}^v)}\big),
\end{equation}
where $\dot{x}_{\mathrm{r},i}^v(t) \!=\! 0$ with $\sigma_i \!=\! 0$, $\boldsymbol{u}_i(t) \!=\! \boldsymbol{r}_i \, x_{\mathrm{r},0,i}^v \!=\! \textrm{const}$ and $x_{\mathrm{r},i}^v(t) \!=\! x_{\mathrm{r},0,i}^v \!=\! \textrm{const}$ hold for $i=1,\ldots,r$. In other words, the employment of a zero signal generator corresponds to moment matching at shifts $\sigma_i \!=\! 0$. 

\subsection{Time discretization with collocation points}
% except for zero signal generator, the equations () and () are time/state-dependent, cannot be simplified as it is done for linear systems...
% Therefore, time collocation
Except for the case with a zero signal generator, the nonlinear equations \eqref{eq:LP-NSG-elem} and \eqref{eq:LP-LSG-elem} are state-dependent and cannot be solved so easily. Remember that in the linear case, the state vector $\boldsymbol{x}_{\mathrm{r}}^v(t)$ could be factored out, yielding a constant linear matrix equation that is satisfied for $\boldsymbol{x}_{\mathrm{r}}^v(t)$. Unfortunately, this factorization cannot be generally done in the non\-li\-near setting anymore. Thus, inspired by POD, we propose to discretize the state-dependent equations with \emph{time-snapshots} or \emph{collocation points} $\left\{t^*_k\right\}$, $k=1,\ldots,K$.

\subsubsection{Nonlinear signal generator} \label{subsubsec:LP-NSG-elem-timeDis}
For a time-discretized nonlinear signal generator $s_{v_{i}}(x_{\mathrm{r},i}^v(t^*_k))$, $\boldsymbol{r}_{_i}(x_{\mathrm{r},i}^v(t^*_k))$ and $x_{\mathrm{r},0,i}^v$, the following time-independent equation results 
\begin{equation} \label{eq:LP-NSG-elem-timeDis}
	\boldsymbol{0} = \boldsymbol{f}\big(\boldsymbol{v}_{ik} \, x_{\mathrm{r}, i}^v(t^*_k), \, \boldsymbol{r}_{_i}(x_{\mathrm{r},i}^v(t^*_k))\big) - \boldsymbol{v}_{ik} \, s_{v_{i}}\big(x_{\mathrm{r},i}^v(t^*_k)\big),
\end{equation}
which can be solved for each $\boldsymbol{v}_{ik} \in \mathbb{R}^n$, with $i=1,\ldots,r$ and $k=1,\ldots,K$, if desired. Note that the discrete solution $x_{\mathrm{r},i}^v(t^*_k)$ of the nonlinear signal generator ODE must be given or computed via simulation before solving equation \eqref{eq:LP-NSG-elem-timeDis}. 

\subsubsection{Linear signal generator}
Using the time-discretized signal generator $\dot{x}_{\mathrm{r},i}^v(t^*_k) \!=\! \sigma_i \, x_{\mathrm{r},i}^v(t^*_k)$, $\boldsymbol{u}_i(t^*_k) \!=\! \boldsymbol{r}_i \, x_{\mathrm{r},i}^v(t^*_k)$ and $x_{\mathrm{r},0,i}^v$, equation \eqref{eq:LP-LSG-elem} becomes time-independent
\begin{equation} \label{eq:LP-LSG-elem-timeDis}
	\begin{aligned}
		\boldsymbol{0} &= \boldsymbol{f}\big(\boldsymbol{v}_{ik} \, x_{\mathrm{r},i}^v(t^*_k), \, \boldsymbol{r}_i \, x_{\mathrm{r},i}^v(t^*_k)\big) - \boldsymbol{v}_{ik} \, \sigma_i \, x_{\mathrm{r},i}^v(t^*_k), %\\[0.3em]
		%&= \boldsymbol{f}\big(\boldsymbol{v}_i \, \mathrm{e}^{\sigma_i t_k} x_{\mathrm{r},0,i}, \boldsymbol{r}_i \, \mathrm{e}^{\sigma_i t_k} x_{\mathrm{r},0,i}\big) - \sigma_i \, \boldsymbol{v}_i \, \mathrm{e}^{\sigma_i t_k} x_{\mathrm{r},0,i}
	\end{aligned}
\end{equation}
with $x_{\mathrm{r},i}^v(t^*_k) \!=\! \mathrm{e}^{\sigma_i t^*_k} \, x_{\mathrm{r},0,i}^v$ for $i=1,\ldots,r$. Note that in this case, the discrete solution $x_{\mathrm{r},i}^v(t^*_k)$ of the linear signal generator ODE is analytically given by exponential functions with exponents $\sigma_i$, so that no simulation of the signal generator is required.   

\subsubsection{Zero signal generator} For this special case, no time discretization is needed, since \eqref{eq:LP-ZSG-elem} already represents a time-independent equation. Please note that solving the nonlinear system of equations \eqref{eq:LP-ZSG-elem} is strong related to computing the \emph{steady-state} $\boldsymbol{x}_{\infty}$, also called \emph{equilibrium point}, of the nonlinear system \eqref{eq:nonlin-FOM} by means of $\boldsymbol{0} \!=\! \boldsymbol{f}\big(\boldsymbol{x}_{\infty}, \boldsymbol{u}_{\mathrm{const}}\big)$.

\subsection{Simulation-free nonlinear moment matching algorithm}
After the step-by-step simplifications discussed in the previous section, we are now ready to state our proposed simulation-free nonlinear moment matching algorithm:  
%\renewcommand{\thealgorithm}{}
%\begin{algorithm}[ht]\caption{Nonlinear Moment Matching (NLMM)} \label{alg:nlmm} 
\begin{algorithm}[!ht]\caption{Nonlinear Moment Matching (NLMM)} \label{alg:nlmm}
	\begin{algorithmic}[1]
		\Require $\boldsymbol{f}(\boldsymbol{x}, \!\boldsymbol{u})$, $\!\boldsymbol{J}_{\boldsymbol{f}}(\boldsymbol{x}, \!\boldsymbol{u})$, $\!s_{v_i}(x_{\mathrm{r},i}^v(t^*_k))$, $\!\boldsymbol{r}_{_i}(x_{\mathrm{r},i}^v(t^*_k))$, $\!x_{\mathrm{r},i}^v(t^*_k)$, \hspace{1em} initial guesses $\boldsymbol{v}_{0,ik}$, deflated reduced order $r_{\mathrm{defl}}$ \vspace{0.2em}
		\Ensure orthogonal basis $\boldsymbol{V}$ \vspace{0.2em}
		\For{\begin{small} \texttt{i = 1 : r} \end{small}} \hspace{1.5em} $\triangleright$ e.g. $r$ different shifts $\sigma_i$
		\For{\begin{small} \texttt{k = 1 : K} \end{small}} \vspace{0.2em} \hspace{0.4em} $\triangleright$ e.g. $K$ samples in each shift 
		\State \begin{small} \hspace{-1.5em} \texttt{fun=@(v)} $\boldsymbol{f}\big(\texttt{v*}x_{\mathrm{r},ik}^v, \, \boldsymbol{r}_{_i}(x_{\mathrm{r},ik}^v)\big)  - \texttt{v*}s_{v_i}(x_{\mathrm{r},ik}^v)$ \end{small} \label{al:line:fun} \vspace{0.3em}
		\State \begin{footnotesize} \hspace{-1.5em} \texttt{Jfun=@(v)} $\boldsymbol{J}_{\boldsymbol{f}}\big(\texttt{v*}x_{\mathrm{r},ik}^v, \, \boldsymbol{r}_{_i}(x_{\mathrm{r},ik}^v)\big)\texttt{*}x_{\mathrm{r},ik}^v - \boldsymbol{\mathrm{I}}_n\texttt{*}s_{v_i}(x_{\mathrm{r},ik}^v)$ \end{footnotesize} \label{al:line:Jfun} \vspace*{-0.8em}
		\State \begin{small} \hspace{-1.5em} \texttt{V(:,(i-1)*K+k)=} \textbf{\texttt{Newton}}\texttt{(fun,}$\, \boldsymbol{v}_{0,ik} \, $\texttt{,Jfun)} \end{small} \label{al:line:Newton} \vspace{0.3em}
		\State \begin{small} \hspace{-1.4em} \texttt{V = }\textbf{\texttt{gramSchmidt}}\texttt{((i-1)*K+k, V)} \end{small} \label{al:line:gramSchmidt} \vspace{0.2em} \hspace{0.1em} $\triangleright$ optional  %\vspace{0.1em}
		\EndFor
		\EndFor
		\State \texttt{V = }\textbf{\texttt{svd}}\texttt{(V,}$\, r_{\mathrm{defl}}$\texttt{)} \hspace{1.1em} $\triangleright$ deflation is optional \label{al:line:SVD}
	\end{algorithmic}
\end{algorithm}\\
Note that the algorithm is given for the most general case of a nonlinear signal generator (cf. eq. \eqref{eq:LP-NSG-elem-timeDis}), and where \emph{two} nested \textbf{for}-loops are used to compute all possible $\boldsymbol{v}_{ik} \in \mathbb{R}^n$. Nevertheless, other (simpler) strategies are also conceivable. These and further aspects are discussed in the following.

\paragraph{Different strategies and degrees of freedom} 
In a\-ddi\-tion to a nonlinear signal generator, one could also apply a linear or a zero signal generator. To this end, line \ref{al:line:fun} (and correspondingly line \ref{al:line:Jfun} also) in Algorithm \ref{alg:nlmm} should be replaced according to the equations \eqref{eq:LP-LSG-elem-timeDis} and \eqref{eq:LP-ZSG-elem}. Note again that the latter cases do not require the simulation of the signal generator ODE to compute $x_{\mathrm{r},i}^v(t^*_k)$. Moreover, please remember the importance of the choice of an adequate signal generator for a suitable characterization and reduction of the nonlinear system at hand.  
%Moreover, please remember the importance of the chosen signal generator for the characterization and reduction quality of the nonlinear system at hand.

Besides the depicted most general approach, where basis vectors are computed for different signal generators~$i\!=\!1,\ldots,r$ at several collocation points $k \!=\! 1, \ldots,K$, one could also consider some special cases. For instance, a single signal generator ($r\!=\!1$) at several collocation points $k \!=\! 1, \ldots,K$ is a possible simpler approach. Herein, the choice of appropriate time-snapshots $t^*_k$ of the selected signal generator is of crucial importance. Another procedure consists in matching moments for different signal generators $i \!=\! 1,\ldots,r$ at only one time-snapshot ($K \!=\! 1$). This multipoint moment matching strategy implies, exemplarily for a linear signal generator, the choice of different shifts and tangential directions $\left\{\sigma_i, \boldsymbol{r}_i\right\}$, which may be selected e.g. logarithmically between $\left[\omega_{\mathrm{min}}, \omega_{\mathrm{max}}\right]$ or via IRKA \cite{gugercin2008h_2}.
For a zero signal generator this implies the choice of different initial conditions and tangential directions $\left\{x_{\mathrm{r},0,i}^v, \boldsymbol{r}_i\right\}$.

\paragraph{Computational effort}
The presented reduction technique is \emph{simulation-free}, since it does not require the numerical integration of the large-scale nonlinear system \eqref{eq:nonlin-FOM}. However, it involves the solution of (at most $r \cdot K$) nonlinear systems of equations (NLSE) of full order dimension $n$. These NLSEs can be solved using either a self-programmed Newton-Raphson scheme (cf. line \ref{al:line:Newton}) or the \textsc{MATLAB}'s built-in function \textbf{\texttt{fsolve}}. For a faster convergence of the Newton method, it is highly recommended to supply the analytical Jacobian of the right-hand side \texttt{Jfun}, for which the Jacobian of the nonlinear system $\boldsymbol{J}_{\boldsymbol{f}}(\boldsymbol{x}, \boldsymbol{u})$ is needed. If \texttt{Jfun} is not provided, then the Jacobian is approximated using finite differences, which can be very time-consuming.  
Further note that reduction techniques like POD require a, typically implicit, numerical simulation of the FOM, which also relies on the solution of NLSEs with the Newton-Raphson method. However, the computational effort of a forward simulation compared to NLMM is supposed to be higher, since -- within a simulation -- a NLSE must be solved in \emph{each} time-step.  

\paragraph{Other aspects}
A good initial guess for the solution of a NLSE can considerably speed-up the convergence of the Newton method. Towards this aim, initial guesses can be taken from linearized models, i.e. $\boldsymbol{v}_{0,i} \!=\! (\sigma_i \boldsymbol{\mathrm{I}} \!-\! \boldsymbol{A})^{-1}\boldsymbol{B} \boldsymbol{r}_i$, or from the solutions at neighbouring shifts or time-snapshots. 

Another important aspect is that the matrix $\boldsymbol{V}$ containing all basis vectors $\boldsymbol{v}_{ik}$ must have full rank, and should pre\-fe\-ra\-bly be orthogonal for better numerical robustness. Thus, if too many or redundant columns are available, a deflation can be performed (cf. line \ref{al:line:SVD}). Moreover, a Gram-Schmidt orthogonalization process can optionally be employed.
\vspace{-0.2em}
\subsection{Analysis and Discussion}
	In this section, a discussion about the proposed simplifications and the presented simulation-free algorithm is given.
	Firstly, it is important to note that the use of a linear projection resembles a special case of the most general nonlinear projection framework, or the polynomial expansion-based ansatz proposed in \cite{krener1992construction,huang2004nonlinear} and used in \cite{scarciotti2017data}. In fact, applying a more sophisticated projection ansatz with basis functions customized for the nonlinear system at hand seems promising for future research. Interestingly, this polynomial ansatz seems to be also linked to the Volterra series representation often used for the reduction of special nonlinear system classes \cite{rugh1981nonlinear,breiten2013interpolatory,cruz2018nonlinear}. Nevertheless, a linear projection might be sufficient in certain cases and its use is motivated here by its simplicity and its frequent and successful employment in nonlinear MOR.    
	Secondly, it is emphasized again that the choice of the signal generator is crucial for the quality of the reduced model. Hence, it should be selected according to the nonlinear system to be reduced. The validity of the special linear signal generator for characterizing nonlinear systems is questionable and not completely clear yet, but it has been shown that this type of signal generator (together with a linear projection) is being implicitly applied also for the reduction of bilinear and quadratic bilinear systems \cite{cruz2018nonlinear}. 

%%%%%%%%%%%%%%%%%%%%%%%%%%%%%%%%%%%%%%%%%%%%%%%%%%%%%%%%%%%%%%%%%%%%%%%%%%%%%%%%
\section{Numerical Example}
The efficiency of the proposed simulation-free nonlinear moment matching algorithm is illustrated by means of the FitzHugh-Nagumo (FHN) benchmark model from \cite{chaturantabut2010nonlinear}. This model describes the activation and deactivation dynamics of a spiking neuron. A spatial discretization of the underlying coupled nonlinear PDE into $\ell$ elements yields a model of $n \!\!=\!\! 2 \ell$ degrees of freedom. The model equation is given by
	\vspace{-0.2em}
	\begin{equation} \label{eq:FHN}
		\begin{aligned}
			\boldsymbol{E} \, \dot{\boldsymbol{x}}(t) &= \overbrace{\boldsymbol{A} \, \boldsymbol{x}(t) + \boldsymbol{\tilde{f}}\big(\boldsymbol{x}(t)\big) + \boldsymbol{B} \, \boldsymbol{u}(t)}^{\boldsymbol{f}\left(\boldsymbol{x}(t), \boldsymbol{u}(t)\right)}, \\
			\boldsymbol{y}(t) &= \underbrace{\boldsymbol{C} \, \boldsymbol{x}(t)}_{\boldsymbol{h}\left(\boldsymbol{x}(t)\right)},
		\end{aligned}
	\end{equation}
	%\vspace{-0.7em}
	with a cubic nonlinea\-ri\-ty $\tilde{f}(v_\ell) \!=\! v_\ell(v_\ell - 0.1)(1 - v_\ell)$ and $\boldsymbol{x} \!\!=\!\! \left[\boldsymbol{v}^{\mathsf T}, \boldsymbol{w}^{\mathsf T}\right]^{\mathsf T}$. The state variables $v_\ell$ and $w_\ell$ represent the voltage and recovery voltage at each spatial element. The model is input-affine with $\boldsymbol{u}(t) \!=\! \left[i_0(t), \, 1\right]^{\mathsf T}$, where $i_0(t) \!=\! 5 \cdot 10^4 \, t^3 \, \mathrm{e}^{-15 t}$ denotes the electric current excitation. The outputs are chosen at the left boundary ($z\!=\!0$) via the output matrix $\boldsymbol{C}$, i.e. $\boldsymbol{y}(t) \!=\! \left[v_1(t), w_1(t)\right]^{\mathsf T}$. Please note that $\boldsymbol{E} \neq \boldsymbol{\mathrm{I}}$. Since the matrix $\boldsymbol{E}$ in this example is however diagonal, it can be efficiently carried to the right-hand side by its inverse $\boldsymbol{E}^{-1}$ to obtain the explicit representation \eqref{eq:nonlin-FOM} with $\boldsymbol{E} = \boldsymbol{\mathrm{I}}$. Note that, for systems with more general, regular $\boldsymbol{E}$, it is advisable to apply the reduction directly to the implicit state-space representation instead of using the inverse. To this end, Algorithm \ref{alg:nlmm} can be extended in a straightforward manner.

For the application of Algorithm \ref{alg:nlmm}, in this case a single signal generator ($r \!=\! 1$) with $K \!=\! 41$ equidistant time-snapshots in the interval $\left[0, 5\right]$ is considered. The following linear signal generator $\dot{x}_{\mathrm{r}}^v(t) \!=\! x_{\mathrm{r}}^v(t) + 0.3$, $\boldsymbol{u}(t) \!=\! \left[x_{\mathrm{r}}^v(t), \, 1\right]^{\mathsf T}$, $x_{\mathrm{r},0}^v \!=\! -0.29$ is chosen, since the solution of the ODE is given by $x_{\mathrm{r}}^v(t) \!=\! \mathrm{e}^{t} \, x_{\mathrm{r},0}^v + 0.3 (\mathrm{e}^{t} - 1)$. Hence, this signal represents a growing exponential function shifted along the negative $y$-axis, whose values cover the interesting value range $\left[-0.29, 1.18\right]$ of the state variables. Please note that this unstable input signal is only used to compute the projection matrix $\boldsymbol{V}_{\text{NLMM}}$ via Algorithm \ref{alg:nlmm} during the \emph{training phase}. For the \emph{test phase}, the above input with the current $i_0(t)$ has been applied. Regarding POD, the input $i_0(t)$ has been applied for \emph{both} the training and test phase. This means that $\boldsymbol{V}_{\text{POD}}$ has been constructed and tested with the very same input signal. This rather unfair scenario has been chosen intentionally to assess the potential of NLMM. The numerical results of this scenario are quantitatively summarized in Table \ref{tab:results} and exemplarily illustrated for $r_{\mathrm{defl}} \!=\! 22$ in Fig. \ref{fig:num-results}.

%\begin{table}[h] 
%	\centering
%	\caption{Numerical results and comparison between POD and NLMM}
%	\begin{tabular}{c c c c}
%		\toprule
%		FHN ($k \!=\! 100 $) &  red. time & sim. time & rel. $\mathcal{L}_1$ error norm  \\
%		\midrule
%		FOM ($n \!=\! 200 $) & - & 1000 & -\\[0.3em]
%		POD ($r_{\mathrm{defl}} \!=\! 22$) & 21 & 46546 & 7575 \\[0.3em]
%		NLMM ($r_{\mathrm{defl}} \!=\! 22$) & 34 & 4 & 32 \\[0.1em]
%	\end{tabular}
%	
%	\vspace{0.2em}
%	\begin{tabular}{c c c c}
%		\toprule \toprule
%		FHN ($k \!=\! 1000 $) &  red. time & sim. time & rel. $\mathcal{L}_1$ error norm  \\
%		\midrule
%		FOM ($n \!=\! 2000 $) & - & 1000 & -\\[0.3em]
%		POD ($r_{\mathrm{defl}} \!=\! 34$) & 21 & 575 & 5775 \\[0.3em]
%		NLMM ($r_{\mathrm{defl}} \!=\! 34$) & 34 & 4 & 32 \\
%		\bottomrule
%	\end{tabular}
%	\label{tab:results1}
%\end{table} 
%\vspace{-0.2em}
\begin{table}[h] 
	\centering
	\caption{Numerical comparison between POD and NLMM}
	\begin{tabular}{c c c c}
		\toprule
		FHN ($\ell \!=\! 1000 $) &  red. time & sim. time & rel. $\mathcal{L}_1$ error norm  \\
		\midrule
		FOM ($n \!=\! 2000 $) & - & \unit[382.16]{s} & - \\
		\midrule \midrule
		POD ($r_{\mathrm{defl}} \!=\! 22$) & \unit[382.25]{s} & \unit[28.29]{s} & $1.03 \, \mathrm{e}^{-5}$  \\[0.3em]
		NLMM ($r_{\mathrm{defl}} \!=\! 22$) & \unit[46.17]{s} & \unit[28.91]{s} & $3.36 \, \mathrm{e}^{-3}$ \\[0.1em]
		\midrule \midrule
		POD ($r_{\mathrm{defl}} \!=\! 34$) & \unit[382.26]{s} & \unit[31.84]{s} & $2.17 \, \mathrm{e}^{-8}$ \\[0.3em]
		NLMM ($r_{\mathrm{defl}} \!=\! 34$) & \unit[47.23]{s} & \unit[30.86]{s} & $1.83 \, \mathrm{e}^{-3}$ \\[0.1em]
		\bottomrule
	\end{tabular}
	\label{tab:results}
\end{table}
\vspace{-1.3em}
\begin{figure}[h!]
	\begin{center} 
		\ref{named}\\[-0.1em]
		\setlength\mywidth{0.18\textwidth}
		\setlength\myheight{0.8\mywidth} %0.61803
		\subfloat{\centering
			\input{./matlab/limitCyclen2000r22.tikz}		
			\label{fig:limitCyclen2000r22}}
		\subfloat{\centering
			\input{./matlab/outputsn2000r22.tikz}
			\label{fig:outputsn2000r22}} %\\
		%		\ref{named}	
	\end{center}
	\vspace{-1em}
	\caption{\footnotesize Limit cycle behavior and outputs of the FHN model for test signal $\boldsymbol{u}(t) \!=\! \left[i_0(t), \, 1\right]^{\mathsf T}$ with $i_0(t) \!=\! 5 \cdot 10^4 \, t^3 \, \mathrm{e}^{-15 t}$ \  ($r_{\mathrm{defl}} \!=\! 22$)}
	\label{fig:num-results}
\end{figure}
\vspace{-0.7em}
%
%	\begin{center}% note that \centering uses less vspace...
%		\subfloat[Anfangswertterm~$(\cdot)_{\mathrm{AW}}$]{\centering
%			\input{images/matlab/blin_AW2.tex}		
%			\label{fig:blingAW}}%	
%		\subfloat[Anregungsterm~$(\cdot)_{\mat B}$]{\centering
%			\input{images/matlab/blin_B2.tex}		
%			\label{fig:blingB}}\\[1ex]	
%		\ref{named}
%	\end{center}
%
%
The comparison between POD and NLMM in terms of computational effort shows that NLMM requires less time to compute the deflated basis $\boldsymbol{V}$ than POD. Note that the latter relies first on the training simulation of the FOM (using $i_0(t)$) within an \textbf{\texttt{implicitEuler}} scheme with the fixed-step size $h \!=\! \unit[0.01]{s}$, and then on a singular value decomposition (SVD) of the gained snapshot matrix. By contrast, NLMM needs to solve $K \!=\! 41$ NLSEs (using the unstable signal generator) and to perform an SVD of a smaller matrix. \\
In terms of approximation quality, both approaches yield satisfactory numerical results with moderate relative error norms between FOM and ROM using $i_0(t)$ as test signal, even though for NLMM a growing exponential input has been applied during the training phase.

%%%%%%%%%%%%%%%%%%%%%%%%%%%%%%%%%%%%%%%%%%%%%%%%%%%%%%%%%%%%%%%%%%%%%%%%%%%%%%%%
\section{CONCLUSIONS}
In this contribution, the concept of moment matching known from linear systems is first revisited and then comprehensively explained for nonlinear systems based on \cite{astolfi2010model}. Then, some simplifications are proposed, yielding a ready-to-implement, simulation-free nonlinear moment matching algorithm, which relies on the solution of NLSEs rather than of a PDE. Hereby, some useful theoretical insights concerning the meaning and the importance of the chosen signal generator are provided, and the diverse strategies and numerical aspects of the proposed algorithm are discussed. All in all, it can be concluded that the signal generator, i.e. the chosen input, plays a crucial role and should characterize the nonlinear system at hand.

%%%%%%%%%%%%%%%%%%%%%%%%%%%%%%%%%%%%%%%%%%%%%%%%%%%%%%%%%%%%%%%%%%%%%%%%%%%%%%%%%
%\section*{ACKNOWLEDGMENTS}
%
%The authors gratefully acknowledge the reviewers' comments.
%The authors would like to thank the reviewers for the constructive comments on how to improve the manuscript.

%%%%%%%%%%%%%%%%%%%%%%%%%%%%%%%%%%%%%%%%%%%%%%%%%%%%%%%%%%%%%%%%%%%%%%%%%%%%%%%%
%\cite{antoulas2005approximation}
%\cite{astolfi2010model}
%\cite{astolfi2010steady} %duality
%\cite{ionescu2013families} % duality
%\cite{ionescu2016nonlinear} %nonlinear projection, duality
%\cite{scarciotti2017data}
%\cite{scarciotti2017review}
%\cite{isidori1995nonlinear}

%\cite{beattie2017model}
%\cite{benner2017model}
%\cite{gallivan2004sylvester}
%\cite{grimme1997krylov}

%\cite{baur2014model} % system-theoretic approaches
%\cite{moore1981principal} % Pod
%\cite{kunisch1999control} % Pod
%\cite{kunisch2008proper} % Pod
%\cite{willcox2002balanced} % Balanced-POD, Empirical Gramians
%\cite{rewienski2003trajectory} % Tpwl
%\cite{lall2002subspace} % Empirical Gramians
%\cite{breiten2013interpolatory}
%\cite{rugh1981nonlinear}
%\cite{scherpen1993balancing}
\vspace{-0.2em}
\bibliographystyle{abbrv}
\bibliography{NLMOR_NLMM_ECC19}

\end{document}

%% file: auxfiles/blockschaltbilder.tex
\newcommand{\Summationsstelle}[4][]{ %
  \begin{scope}[thick, #1]
    \node[draw, circle, minimum width = #4, minimum height = #4] (#2) at (#3) {};
  \end{scope} %
}

\newcommand{\Verzweigung}[4][]{ %
  \begin{scope}[inner sep = 0 pt, outer sep = 0 pt, thick, #1]
    \node[circle, minimum width = #4, minimum height = #4] (#2) at (#3) {};
    \draw[solid, fill] (#2) circle (0.5*#4);
  \end{scope} %
}

\newcommand{\UeFunk}[5][]{ %
  \begin{scope}[thick, #1]
    \node[minimum width = #4, minimum height = #4] (#2) at (#3) {#5};
    \draw (#2.north east) rectangle (#2.south west);
  \end{scope} %
}

%% file: tikz/lin-sys-lin-SG_steady-state-V.tex
\begin{tikzpicture}

  % Knoten / Blöcke
  \UeFunk[align = left, inner sep = 6 pt]{linSigGen}{-0.8, 0}{0.6 cm}{
		$\begin{aligned}
		\dot{\boldsymbol{x}}_{\mathrm{r}}^v(t) &= \boldsymbol{S}_{v} \, \boldsymbol{x}_{\mathrm{r}}^v(t) \\[0.2em]
		\boldsymbol{u}(t) &= \boldsymbol{R} \, \boldsymbol{x}_{\mathrm{r}}^v(t)
		\end{aligned}$
  }
  \UeFunk[align = left, inner sep = 6 pt]{FOM}{6.2, 0}{0.6 cm}{
	    $\begin{aligned}
	    \boldsymbol{E} \, \dot{\boldsymbol{x}}(t) &= \boldsymbol{A} \, \boldsymbol{x}(t) + \boldsymbol{B} \, \boldsymbol{u}(t) \\[0.2em]
	    \boldsymbol{y}(t) &= \boldsymbol{C} \, \boldsymbol{x}(t) 
	    \end{aligned}$
  } 
  \UeFunk[align = left, inner sep = 6 pt]{ROM}{6.2, -4}{0.6 cm}{
	  	$\begin{aligned}
     	\boldsymbol{W}^{\mathsf T} \boldsymbol{E} \boldsymbol{V} \, \dot{\boldsymbol{x}}_{\mathrm{r}}(t) &= \boldsymbol{W}^{\mathsf T} \boldsymbol{A} \boldsymbol{V} \, \boldsymbol{x}_{\mathrm{r}}(t) + \boldsymbol{W}^{\mathsf T} \boldsymbol{B} \, \boldsymbol{u}(t) \\[0.2em]
     	\boldsymbol{y}_{\mathrm{r}}(t) &= \boldsymbol{C} \boldsymbol{V} \, \boldsymbol{x}_{\mathrm{r}}(t)
     	\end{aligned}$
  }
  
  \Verzweigung{verzw-u}{2, 0}{3 pt}
  
  \Summationsstelle{summation}{11, 0}{0.4 cm}

  % Signalflüsse
  \draw[thick, -latex'] (linSigGen) -- (FOM) node[pos = 0.49, above] {$\boldsymbol{u}(t) = \boldsymbol{R} \, \mathrm{e}^{\boldsymbol{S}_v t} \, \boldsymbol{x}_{\mathrm{r},0}^v$};
  \draw[thick, -latex'] (verzw-u) |- (ROM);
  \draw[thick, -latex'] (FOM.east) -- (summation) node[pos = 0.51, above] {$\boldsymbol{y}(t)$};
  \draw[thick, -latex'] (ROM.east) -| (summation) node[pos = 0.28, above] {$\boldsymbol{y}_{\mathrm{r}}(t)$};
  \draw[thick, -latex'] (summation) -- + (1.9, 0) node[pos = 0.49, above] {$\boldsymbol{e}(t)=\boldsymbol{0}$};
  
  \draw[thick, -latex'] (linSigGen) -- + (0,1.7) node[pos = 0.7, right] {$\boldsymbol{x}_{\mathrm{r},0}^v \neq \boldsymbol{0}$} -- (linSigGen);

  \draw[thick, -latex'] (FOM) -- + (0,1.7) node[pos = 0.7, right] {$\boldsymbol{x}_{0} = \boldsymbol{V} \boldsymbol{x}_{\mathrm{r},0}^v$} -- (FOM);
  \draw[thick, -latex'] (ROM) -- + (0,1.7) node[pos = 0.7, right] {$\boldsymbol{x}_{\mathrm{r},0} = (\boldsymbol{W}^{\mathsf T} \boldsymbol{E} \boldsymbol{V})^{-1} \boldsymbol{W}^{\mathsf T} \boldsymbol{E} \, \boldsymbol{x}_{0}$} -- (ROM);

  % Beschriftungen
  \node[outer sep = 0.1 cm, below] at (linSigGen.south) {\text{linear signal generator}};
  \node[outer sep = 0.1 cm, below] at (FOM.south) {\text{FOM}};
  \node[outer sep = 0.1 cm, below] at (ROM.south) {\text{ROM}};
  
  \node[outer sep = 0.3 cm, below] at (summation.west) {$-\ $};
  
\end{tikzpicture}

%% file: tikz/nonlin-sys-nonlin-SG_steady-state-V.tex
\begin{tikzpicture}

  % Knoten / Blöcke
  \UeFunk[align = left, inner sep = 6 pt]{nonlinSigGen}{0, 0}{0.6 cm}{
		$\begin{aligned}
		\dot{\boldsymbol{x}}_{\mathrm{r}}^v(t) &= \boldsymbol{s}_{v}\big(\boldsymbol{x}_{\mathrm{r}}^v(t)\big) \\[0.2em]
		\boldsymbol{u}(t) &= \boldsymbol{r}\big(\boldsymbol{x}_{\mathrm{r}}^v(t)\big)
		\end{aligned}$
  }
  \UeFunk[align = left, inner sep = 6 pt]{FOM}{6.7, 0}{0.6 cm}{
	    $\begin{aligned}
	    \dot{\boldsymbol{x}}(t) &= \boldsymbol{f}\big(\boldsymbol{x}(t), \boldsymbol{u}(t)\big) \\[0.2em]
	    \boldsymbol{y}(t) &= \boldsymbol{h}\big(\boldsymbol{x}(t)\big)
	    \end{aligned}$
  } 
  \UeFunk[align = left, inner sep = 6 pt]{ROM}{6.7, -4}{0.6 cm}{
	  	$\begin{aligned}
     	\dot{\boldsymbol{x}}_{\mathrm{r}}(t) &= \left.\frac{\partial \boldsymbol{\omega}(\boldsymbol{x})}{\partial \boldsymbol{x}}\right|_{\boldsymbol{x}=\boldsymbol{\nu}(\boldsymbol{x}_{\mathrm{r}})} \, \boldsymbol{f}\big(\boldsymbol{\nu}(\boldsymbol{x}_{\mathrm{r}}(t)), \boldsymbol{u}(t)\big) \\[0.2em]
     	\boldsymbol{y}_{\mathrm{r}}(t) &= \boldsymbol{h}\big(\boldsymbol{\nu}(\boldsymbol{x}_{\mathrm{r}}(t))\big)
     	\end{aligned}$
  }
  
  \Verzweigung{verzw-u}{2.5, 0}{3 pt}
  
  \Summationsstelle{summation}{11.5, 0}{0.4 cm}

  % Signalflüsse
  \draw[thick, -latex'] (nonlinSigGen) -- (FOM) node[pos = 0.49, above] {$\boldsymbol{u}(t)$};
  \draw[thick, -latex'] (verzw-u) |- (ROM);
  \draw[thick, -latex'] (FOM.east) -- (summation) node[pos = 0.76, above] {$\boldsymbol{y}(t)$};
  \draw[thick, -latex'] (ROM.east) -| (summation) node[pos = 0.25, above] {$\boldsymbol{y}_{\mathrm{r}}(t)$};
  \draw[thick, -latex'] (summation) -- (13.4, 0) node[pos = 0.49, above] {$\boldsymbol{e}(t)=\boldsymbol{0}$};
  
  \draw[thick, -latex'] (nonlinSigGen) -- + (0,1.7) node[pos = 0.7, right] {$\boldsymbol{x}_{\mathrm{r},0}^v \neq \boldsymbol{0}$} -- (nonlinSigGen);

  \draw[thick, -latex'] (FOM) -- + (0,1.7) node[pos = 0.7, right] {$\boldsymbol{x}_{0} = \boldsymbol{\nu}(\boldsymbol{x}_{\mathrm{r},0}^v)$} -- (FOM);
  
  \draw[thick, -latex'] (ROM) -- + (0,1.7) node[pos = 0.7, right] {$\boldsymbol{x}_{\mathrm{r},0} = \boldsymbol{\omega}(\boldsymbol{x}_{0})$} -- (ROM);

  % Beschriftungen
  \node[outer sep = 0.1 cm, below] at (nonlinSigGen.south) {\text{nonlinear signal generator}};
  \node[outer sep = 0.1 cm, below] at (FOM.south) {\text{FOM}};
  \node[outer sep = 0.1 cm, below] at (ROM.south) {\text{ROM}};
  
  \node[outer sep = 0.3 cm, below] at (summation.west) {$-\ $};
  
\end{tikzpicture}